\documentclass[conference,a4paper]{IEEEtran}        % 使用 IEEEtran 文档类，适用于会议论文，纸张大小为 A4
\IEEEoverridecommandlockouts                        % 覆盖 IEEEtran 类中的某些命令锁定状态，允许自定义命令
\usepackage[a4paper, top=20mm, bottom=30mm, left=18mm, right=18mm]{geometry}  % 用于调整页边距的包，当前被注释掉
\usepackage[hidelinks]{hyperref}                    % 提供超链接功能，hidelinks 选项隐藏超链接的边框
\usepackage[cmex10]{amsmath}                        % 提供高级数学公式排版功能，cmex10 选项指定数学扩展字体
\usepackage{amssymb,amsfonts}                       % 提供额外的数学符号和字体支持
\usepackage{dblfloatfix}                            % 修复双栏模式下的浮动体（如图表）排版问题
\usepackage[ruled,vlined]{algorithm2e}              % 用于排版算法，ruled 和 vlined 是算法的样式选项
\usepackage{graphicx}                               % 提供插入图形的功能
\usepackage{booktabs}                               % 提供高质量的表格排版功能
\usepackage{siunitx}                                % 用于排版单位和数字
\usepackage[numbers,compress]{natbib}               % 提供参考文献排版功能，numbers 和 compress 是参考文献样式选项
\usepackage{texnames}                               % 用于处理 TeX 相关的名称
\usepackage{bm,bbm}                                 % 提供数学符号和字体的支持
\usepackage{orcidlink}                              % 用于插入 ORCID 链接
\usepackage{tabularx}                               % 提供自动调整宽度的表格功能
\usepackage{makecell}                               % 提供复杂的表格单元格排版功能
\usepackage{multirow}                               % 提供跨行表格功能
\usepackage{array}                                  % 增强表格功能
\usepackage{upgreek}                                % 提供希腊字母排版功能
\usepackage{tablefootnote}                          % 提供在表格中添加脚注的功能
\usepackage{threeparttable}                         % 提供带有注释的表格功能
\usepackage{hyperref}       % 用于插入超链接

\begin{document}
\topmargin=0mm
\title{\uppercase{LCB-CV-UNet: Enhanced Detector for High Dynamic Range Radar Signals}
\thanks{Thanks to the National Natural Science Foundation of China under Grant 62271126 for funding. \\
© [IGARSS2025] IEEE. Personal use of this material is permitted. Permission from IEEE must be obtained for all other uses, in any current or future media, including reprinting/republishing this material for advertising or promotional purposes, creating new collective works, for resale or redistribution to servers or lists, or reuse of any copyrighted component of this work in other works. \\
This work has been accepted for publication in the Proceedings of the IEEE International Geoscience and Remote Sensing Symposium (IGARSS 2025). Doi: 10.1109/IGARSS55030.2025.11243251.\\
Part of the code is available at \href{https://github.com/NGC13009/ComPlex-valued-Lib-2-for-PyTorch.git}{https://github.com/NGC13009/ComPlex-valued-Lib-2-for-PyTorch.git}
}
}
\author{\IEEEauthorblockN{Yanbin Wang$^1$, Xingyu Chen$^1$, Yumiao Wang$^1$, Xiang Wang$^1$, Chuanfei Zang$^1$, Guolong Cui$^{1,*}$, Jiahuan Liu$^2$}
    \IEEEauthorblockA{$^1$\textit{University of Electronic Science and Technology of China}, 611731 Chengdu, China \\
        $^2$\textit{Desay SV Intelligent Transportation Technological Institute Research Co. Ltd.}, 516006 Hui Zhou, China\\
        $^*$Corresponding author: cuiguolong@uestc.edu.cn}
}
\maketitle
\begin{abstract}
    We propose the LCB-CV-UNet to tackle performance degradation caused by High Dynamic Range (HDR) radar signals. Initially, a hardware-efficient, plug-and-play module named Logarithmic Connect Block (LCB) is proposed as a phase coherence preserving solution to address the inherent challenges in handling HDR features. Then, we propose the Dual Hybrid Dataset Construction method to generate a semi-synthetic dataset, approximating typical HDR signal scenarios with adjustable target distributions. Simulation results show about 1\% total detection probability improvement with under 0.9\% computational complexity added compared with the baseline. Furthermore, it excels 5\% over the baseline at the range in 11-13 dB signal-to-noise ratio typical for urban targets. Finally, the real experiment validates the practicality of our model.
\end{abstract}
\section{Introduction}
\begin{figure}[h] % aabbcc
    \centering
    \includegraphics[width=1.\linewidth,page=2]{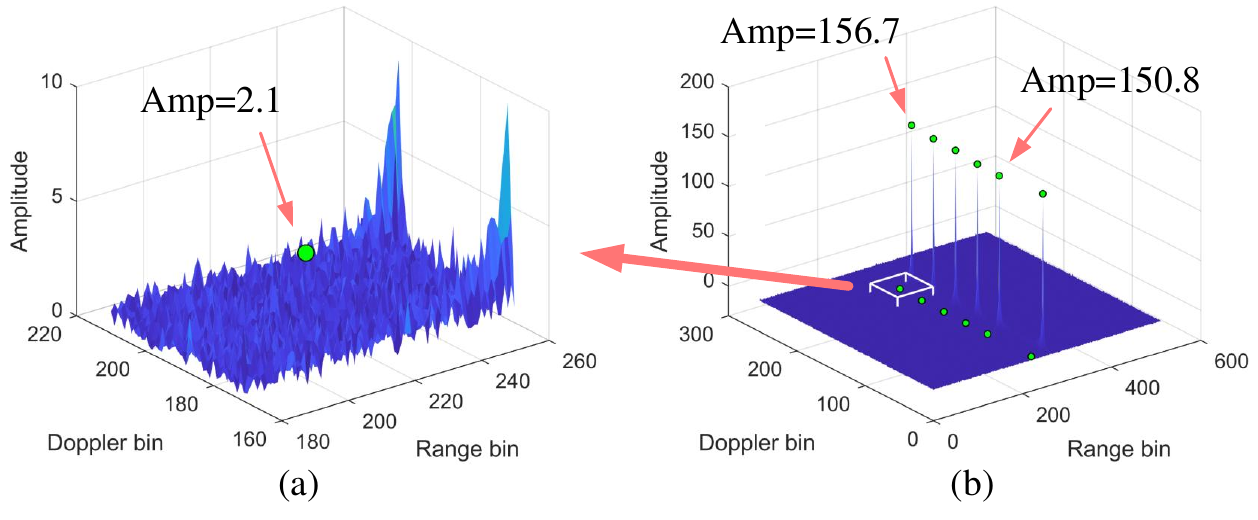}
    \caption{An amplitude illustration of the HDR signal RDM. (a) the magnified region around a weak target with amplitude of $2.1$; (b) global RDM.}
    \label{RDMdemo}
\end{figure}

Deep learning has been extensively applied in multiple fields of radar remote sensing. Nevertheless, radar echo signals surpass RGB images in dynamic range, differing by about two orders of magnitude in amplitude. For example, Fig.~\ref{RDMdemo} depicts an HDR radar signal Range Doppler Map (RDM), where the amplitudes of weak targets are significantly lower than those of strong targets and are obscured by noise. Most vision models are not specifically designed to handle such scenarios. Additionally, in complex environments, strong targets with signal-to-noise ratios (SNR) exceeding $ 20 $dB often produce substantial sidelobes that overlap and obscure low SNR targets. Although filters can suppress sidelobes effectively, this comes at the cost of power loss, thereby degrading the detection performance.

In radar target detection, Constant False Alarm Rate (CFAR) detectors are typically deployed at the edge-side to generate point tracks, reducing the demands on data transmission and storage. However, CFAR detectors often struggle to adapt to strong clutter and interference. Some methods have replaced CFAR with neural networks \cite{pkc}, achieving superior performance, but they fail to address the issue of missed detections for weak targets under HDR signals. Other studies employ complex valued networks \cite{cv_rd_det}, which leverage phase information to enhance weak target detection, but these methods are not always compatible with lightweight platforms, limiting their applicability. Some approaches prioritize lightweight designs to enable edge-side inference \cite{netlite}, but they often underperform when dealing with HDR signals. Other methods focus on constructing datasets to optimize detection performance for specific scenarios \cite{datasetbuild}, yet they neglect the unique characteristics of HDR signals. Neural networks rely on balanced inputs to achieve optimal performance \cite{databalance}. Consequently, training with actual HDR signals will reduce the detection probability for targets with specific SNRs.

In the field of image processing, solutions for handling HDR signals are relatively mature. Multi-exposure fusion is commonly used to preserve details \cite{multifuse}, but it requires multiple spatial and temporal samples for cancellation, which is often impractical to obtain. Tone mapping \cite{sediao1}\cite{sediao2} compresses the dynamic range by adjusting local contrast, but temporal algorithms fail to preserve phase information. Some studies have employed deep learning techniques to tackle HDR challenges \cite{nnhdr}, but these methods typically involve high computational costs, making them unsuitable for edge-side deployment. Logarithmic compression of high intensity signals reduces dynamic range, but it generates unintended anomalous peaks, impairing detection performance of models.

Therefore, this paper introduces a method to enhance radar target detection performance under HDR signals. The key contributions are as follows:
\begin{itemize}
    \item LCB-CV-UNet radar target detection based on Complex-Valued-UNet is proposed to improve detection probability under HDR signals.
    \item The plug-and-play Logarithmic Connect Block (LCB) is proposed, based on the logarithmic connection function, which can computationally efficiently enhance the adaptability of the model to HDR signals.
    \item Dual Hybrid Dataset Construction (DHDC) method is proposed, leveraging HDR signal characteristics, to train neural networks with HDR processing capabilities.
\end{itemize}
Simulations and real datasets evaluations confirm that the model effectively processes HDR signals with minimal computational load.
\section{Methodology}
\begin{figure}[h] % aabbcc
    \centering
    \includegraphics[width=1.\linewidth, page=5]{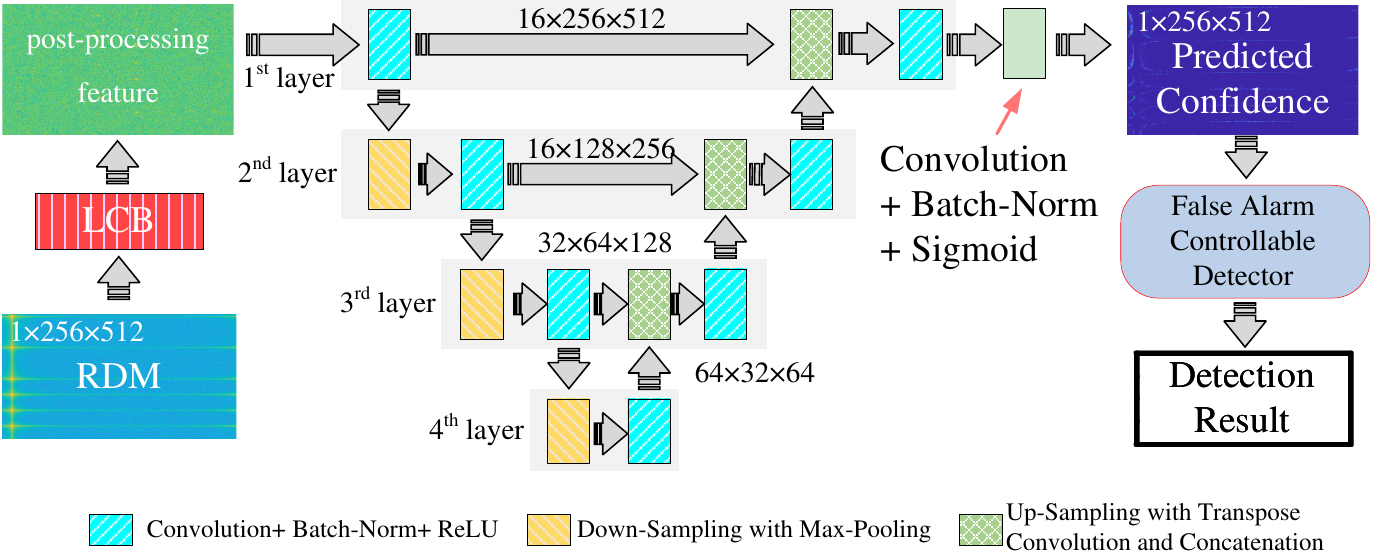}
    \caption{The architecture of the proposed LCB-CV-UNet.}
    \label{LCB_unet}
\end{figure}

The LCB-CV-UNet architecture is depicted in Fig.~\ref{LCB_unet}. It consists of three main components: LCB for preprocessing, CV-UNet for feature extraction, and a CFAR controller as decider. The Up Sampling stage in Fig.~\ref{LCB_unet} includes transposed convolution and copy crop operations. The network takes a single channel radar RDM as input, which is also compatible with other data types. The final probability values for detection units are used to generate detection result point, following the CFAR detection method implemented in the CV-UNet detection network \cite{cvunetw}.
\subsection{Signal Model}
For Linear Frequency Modulated Continuous Wave Multiple Input Multiple Output (LFMCW-MIMO) millimeter wave radar, the received echo is mixed with the transmitted signal, filtering out the high frequency and preserving only the diff-frequency signal to produce the Intermediate Frequency Signal (IFS). Using the Doppler Division Multiple Access (DDMA) scheme, which is widely utilized in modern radar systems, the IFS for the $ q $-th transmitting antenna is denoted as
\begin{align}
    \mathbf{S}_q(m,n) & =A\cdot\exp\left(
    j 2\pi\left( \varphi_{q,m}+f_\text{c}\tau_m+k\tau_m t_{n} \right)
    \right),
    \label{eq:IFS}
\end{align}
where $ \mathbf{S}_q(m,n) $ denotes the element in the $ m $-th row and $ n $-th column of the IFS matrix $ \mathbf{S}_q $. The index $ m $ denotes the pulse number, with $ M $ in total. $ n $ denotes the range cell index, with $ N $ in total. The symbol $A$ denotes the echo amplitude.

In \eqref{eq:IFS}, the first term $ \varphi_{q,m} $ denotes the phase of the transmitted signal. The second term represents the slow-time phase, where $ f_\text{c} $ denotes the carrier frequency. The echo delay denotes $ \tau_m=2(R_0+mVT_\text{p})/c $, where $ R_0 $ denotes the target, $ V $ denotes the target velocity, $ T_\text{p} $ denotes the pulse repetition interval (PRI), and $ c $ denotes the speed of light. The third term represents to the fast-time phase, where $ k $ denotes the chirp slope, and $ t_{n} = n/f_\text{s} $ denotes the fast time, with $ f_\text{s} $ being the sampling frequency. In this paper, $j  = \sqrt{-1}$ is used to denote the imaginary unit.

Next, the IFSs from all transmitting antennas are summed, and pulse compression is performed using the Fast Fourier Transform (FFT) to produce the RDM, which is denoted as
\begin{align}
    \displaystyle \mathbf{X}_{\text{RDM}} = \text{FFT}_\text{2D}\left( \sum_{q}^{}\mathbf{S}_q \right).
\end{align}
The target SNR on the RDM is defined as $ \gamma (\text{dB}) = 10\log_{10}\left( P_\text{S}/P_\text{N} \right) $, where $ P_\text{S} $ is the power of the target cell, and $ P_\text{N} $ is the noise power. If $ \gamma $ spans more than 30 dB, the signal is considered as HDR. Due to the feature of DDMA modulation, a single target will produce multiple peaks.
\subsection{Logarithmic Connect Block}
For a feature element $ x=x_\text{r}+j \cdot x_\text{i}\in\mathbb{C} $, neural networks typically handle the real and imaginary parts separately \cite{cvnet}. However, when processing HDR signals, the extremely large amplitude can cause a substantial dynamic range in both the real and imaginary components, making model fitting more challenging.

To address this issue, we propose a method for processing the complex valued signals derived from HDR features. If $ x_\text{r} $ and $ x_\text{i} $ are handled independently, the condition $ x_\text{i}/x_\text{r}=o_\text{i}/o_\text{r} $ may not be satisfied, where $ o=o_\text{r}+j  o_\text{i} $ denotes the output after processing, leading to phase mismatch noise which can severely impact the performance of detectors. To ensure phase coherence, we adopt the logarithmic connect function, which preserves the phase and facilitates coherent processing, while utilizing its nonlinear properties to enhance the fitting capability of the model. The function defined over $ \mathbb{C} $ is
\begin{align}
    \text{LC}(x) & =
    \begin{cases}
        x,                                                                    & |x| \leq w \\
        \left(w + \ln(1 - w + |x|)\right)\cdot \mathrm{e}^{j  \text{Arg}(x)}, & |x| > w
    \end{cases},
    \label{eq:LC}
\end{align}
where $ \text{Arg}(\cdot) $ denotes the phase angle of a complex number. $w$ denotes the connection position parameter, representing the independent variable value at which two distinct functions are joined. This function reliably preserves the phase relationship and exhibits desirable properties, such as continuity and the existence of continuously differentiable conjugate forms, making it compatible with standard module optimizers.
\begin{algorithm}[h]  % aabbcc
    \label{alg_lcb}
    \DontPrintSemicolon
    \SetKwInput{KwInput}{Input}
    \SetKwInput{KwOutput}{Output}
    \caption{Parallel Computation of LCB}
    \fontsize{8}{10.5}\selectfont
    \KwInput{Signal $\mathbf{X}=[x_1, \cdots, x_N]\in\mathbb{C}^N$, position parameter $w\in\mathbb{R}$.}
    \KwOutput{$\mathbf{O}=[o_1,\cdots,o_N]\in\mathbb{C}^N$.}
    \begin{enumerate}
        \item $\mathbf{R} =\left( \Re(\mathbf{X})^{\circ 2}+\Im(\mathbf{X})^{\circ 2} \right)^{\circ \frac{1}{2}}\in\mathbb{R}^N$
        \item $\mathbf{M}=\mathcal{I}(\mathbf{R}, w)$ // Element operation $\mathcal{I}(x, w)$ = 1 if $x > w$, else $0$     % 此处修改为 \mathcal{I}(\mathbf{R},w)   x>w ....否则计算复杂度就错了
        \item $\mathbf{P}=\mathbf{M}\odot\mathbf{R}+(1-\mathbf{M})\odot w$
        \item $\mathbf{L}=\left(w+\ln(1-w+\mathbf{P})\right)$
        \item $\mathbf{T}=\mathbf{L}\odot\mathbf{M}+\mathbf{R}\odot\left(1-\mathbf{M}\right)$
        \item $\mathbf{S}=\frac{\mathbf{T}}{\mathbf{R}+\varepsilon}$
        \item $\mathbf{O}=\mathbf{S}\odot\mathbf{X}\in\mathbb{C}^N$
        \item Return $\mathbf{O}=[o_1,\cdots,o_N]$
    \end{enumerate}
\end{algorithm}

To achieve lightweight implementation of algorithms on edge-side processors, it is crucial to avoid the use of nonlinear functions, such as Euler's formula. Instead, highly efficient and parallelizable operations should be adopted to fully exploit the parallel acceleration capabilities of these processors, such as FPGAs\cite{fpgagemm}\cite{fpgaenhance}. Accordingly, the LCB algorithm based on $ \text{LC}(\cdot) $ is worked as described in Algorithm~\ref{alg_lcb}. The operator $ \odot $ denotes the Hadamard product, and $\circ a$ denotes the element-wise $a$-th power. The parameter $ \varepsilon > 0 $ is introduced to prevent division by zero errors. The algorithm has a computational complexity of $ \mathcal{O}(N) $ and requires $ 11N $ Multiply-Accumulate Operations (MACs), where one MAC consists of one addition and one multiplication. This is significantly lower than convolution operations based on matrix multiplication. As a result, the algorithm reduces the nonlinear fitting burden on the model, allowing for plug-and-play integration that keeps the network lightweight and improves its ability to coherently process HDR signals. In LCB-CV-UNet, the LCB is embedded at the input stage of the model to tackle the high peak induced by strong targets.
\subsection{Dual Hybrid Dataset Construction Method}
For radar signals in complicated environments, acquiring labeled data is a significant challenge. Moreover, the distribution of target SNR in HDR signals under general conditions may lead to a reduction in detection probability at low SNR levels after model training. Research has proven that training networks with simulated signals is effective \cite{gan}. Therefore, we propose the DHDC method to generate a more balanced dataset, facilitating enhanced training and model performance evaluation. The DHDC method is organized into two methodological parts:
\begin{itemize}
    \item Constructing datasets by superimposing simulated targets into cluttered backgrounds.
    \item Constructing datasets by mixing strong and weak targets according to a controllable distribution.
\end{itemize}

In practical training, to balance the sample distribution, $ \gamma $ of targets over $ w $ are defined as strong targets, where $ \mathcal{P} $ denotes their proportion in the total targets. The distribution is given as follows
\begin{align}
    \gamma\sim \begin{cases}
                   \mathcal{P}\cdot\mathcal{U}(\gamma_{\text{min}},w),     & \gamma<w     \\
                   (1-\mathcal{P})\cdot\mathcal{U}(w,\gamma_{\text{max}}), & \gamma\geq w
               \end{cases}.
\end{align}
In addition, the distance and velocity are sampled from $ R\sim\mathcal{U}(0,R_{\text{max}}) $ and $ V\sim\mathcal{U}(V_{\text{min}},V_{\text{max}}) $, respectively. The number of targets per frame follows $ n\sim\mathcal{U}(1,10) $, where $ \mathcal{U}(a,b) $ denotes a uniform distribution over $ [a,b] $.
\section{Experiment}
To evaluate the effectiveness of the proposed method, we conduct a series of experiments. In the simulation experiments, the DHDC method is used to generate a train dataset of 3000 RDMs and a validation dataset of 1500 RDMs, assuming white Gaussian noise as clutter. The baseline network is the CV-UNet \cite{cvunetw}, tested in the single layer, double layer, and four layer configurations to evaluate the improvement of the LCB under different parameter quantity. Each model is trained under three modes, which can be categorized as follows:
\begin{itemize}
    \item Mix Mode (MM) represents training with a direct mix of strong and non-strong targets with $ \mathcal{P}=0.7 $, serving as the baseline for model performance in HDR signal.
    \item Non-strong Target Mode (NTM) trains exclusively on non-strong targets, representing the detection probability within the particular SNR range of non-strong targets.
    \item Mix+LCB Mode (MM+LCB) uses the proposed LCB-CV-UNet method and the same dataset as MM.
\end{itemize}
All models are trained for 300 epochs, and the lowest-validation-loss model is selected as the final result. Performance is evaluated using the detection probability denoted as $ P_{\text{d}} $ and the average actual false alarm rate denoted as $ P_{\text{fa}} $.

The detection performance is evaluated at a preset false alarm rate of $ \alpha=1\times10^{-4} $. The parameters of the radar system are $ f_\text{c}=77 $ GHz, $ k=29 $ $\text{MHz/}\upmu\text{s}$, $ T_\text{p}=45.6 $ $\upmu\text{s}$, with an ADC sampling rate of $ f_\text{s}=30 $ MHz, $ M=256 $, and $ N=512 $. DDMA modulation is implemented with 6 transmit and 1 receive channels, configured into 8 sub-bands, including $2$ empty bands. The experiments are performed in practical scenarios using the same configuration. For DHDC, the related parameters are set as $R_{\text{max}}=150$ m, $V_{\text{max}}=20$ m/s, $V_{\text{min}}=-20$ m/s, $w=14$, $\gamma_{\text{min}}=6$, and $\gamma_{\text{max}}=60$.
\subsection{Overall Detection Performance}
\begin{figure}[h] % aabbcc
    \centering
    \includegraphics[width=1.\linewidth,page=1]{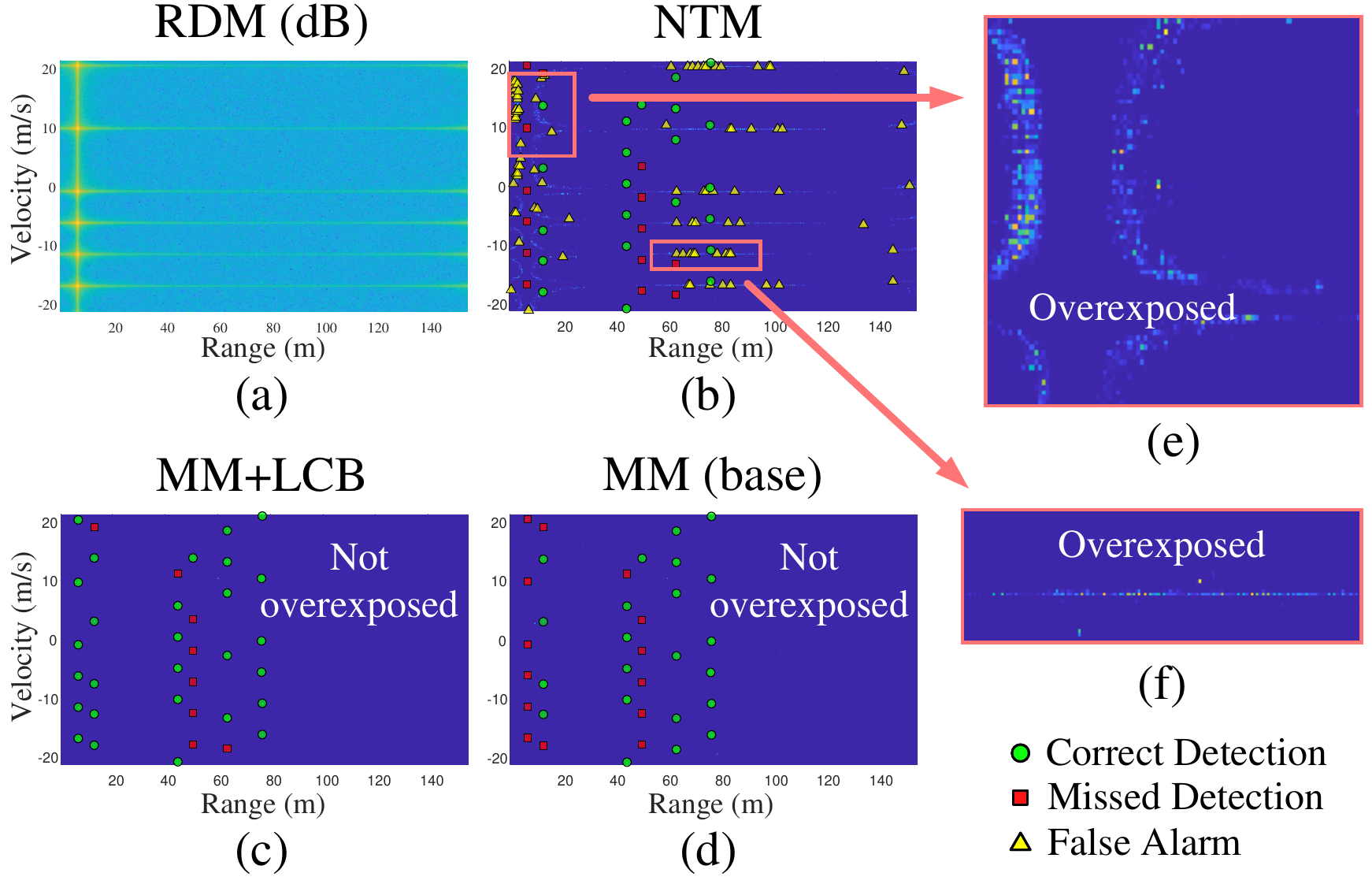}
    \caption{The inference results of the three modes. (a) RDM; (b), (c), and (d) correspond to the results of the three modes; (e) and (f) show magnified views of the indicated regions.}
    \label{wk}
\end{figure}

This experiment evaluates the overall $ P_{\text{d}} $ and the $ P_{\text{fa}} $ under HDR signals, along with an assessment of module complexity. As shown in Table~\ref{mix}, incorporating the LCB increases the overall $ P_{\text{d}} $ by approximately 1\%, while the $ P_{\text{fa}} $ remains within $ (1\pm0.25)\times10^{-4} $, and the computational cost increases by only 0.9\%. Fig.~\ref{wk} shows the case, where (b), (c), and (d) show the results of three modes corresponding to (a) based on $ \alpha=1\times10^{-6} $. Since the NTM is trained solely on non-strong targets, the network becomes overexposed, causing a sharp increase in the false alarm rate and a decline in $ P_{\text{d}} $, as demonstrated by the magnified regions (e) and (f) in Fig.~\ref{wk}.
\vspace{-0.2cm}
\begin{table}[h] % aabbcc
    \renewcommand{\arraystretch}{1.} % 设置行高
    \caption{Results for HDR signal} % 表格标题
    \label{mix} % 表格标签
    \centering % 表格居中
    \begin{tabularx}{\linewidth}{c c l l c c} % 使用 tabularx, 设置总宽度为 \linewidth
        \toprule
        \textbf{Model}$^*$ & \textbf{Method} & \textbf{MACs}     & \textbf{Params}   & $P_{\text{d}} \scriptstyle(\%)$ & \makecell{$P_{\text{fa}}$ \\ ($\scriptstyle\times10^{-4}$)} \\
        \midrule
                           & MM (base)       & 159.908M          & 695B              & 69.24                           & 1.21                      \\
        u1                 & NTM             & 159.908M          & 695B              & 44.69                           & 21.32                     \\
                           & \textbf{MM+LCB} & \textbf{161.349M} & \textbf{695B}     & \textbf{70.21}                  & \textbf{1.23}             \\
        \midrule
                           & MM (base)       & 4.275G            & 23.591K           & 73.24                           & 1.28                      \\
        u2                 & NTM             & 4.275G            & 23.591K           & 47.36                           & 24.29                     \\
                           & \textbf{MM+LCB} & \textbf{4.276G}   & \textbf{23.591K}  & \textbf{74.33}                  & \textbf{1.10}             \\
        \midrule
                           & MM (base)       & 12.478G           & 476.359K          & 73.72                           & 1.19                      \\
        u4                 & NTM             & 12.478G           & 476.359K          & 55.23                           & 13.46                     \\
                           & \textbf{MM+LCB} & \textbf{12.479G}  & \textbf{476.359K} & \textbf{74.36}                  & \textbf{1.08}             \\
        \bottomrule
    \end{tabularx}
    \begin{tablenotes}
        \footnotesize
        \item \hspace{-0.5cm} * u1, u2, and u4 represent the CV-UNet with 1, 2, and 4 layers, respectively.
    \end{tablenotes}
\end{table}
\subsection{Non-strong Target Detection Performance}
In this experiment, the SNR of all targets in the RDM is fixed to evaluate the detection probability for non-strong targets. As shown in Table~\ref{t1s}, for targets with $\gamma$ between 11 and 13 dB, the detection probability of MM+LCB increases by over 5\% compared to MM. This improvement is especially notable for models with lower parameter counts. These results demonstrate that incorporating LCB during training under HDR signals can significantly enhance non-strong targets detection performance. $\bar{P}_{\text{fa}}$ denotes the mean value of $P_\text{fa}$ across all SNR levels.
\begin{table}[h] % aabbcc
    \renewcommand{\arraystretch}{1.} % 设置行高
    \caption{Results for Non-strong SNR target}
    \label{t1s}
    \centering % 表格居中
    \setlength{\tabcolsep}{1.2mm}{
        % \fontsize{7.5}{8.5}\selectfont % 设置字体大小为 9pt, 行距为 11pt
        \begin{tabularx}{\linewidth}{c c c c c c c c c} % 使用 tabularx, 设置总宽度为 \linewidth
            \toprule
            \multirow{2}{*}{\textbf{Model}} & \multirow{2}{*}{\textbf{Method}} & \multicolumn{6}{c}{$P_{\text{d}} \scriptstyle(\%)$ at different SNR levels} & \multirow{2}{*}{\makecell{$\bar{P}_{\text{fa}}$                                                                                     \\ ($\scriptstyle\times10^{-4}$)}} \\ %\cline{3-7}
                                            &                                  & 9dB                                                                         & 10dB                                            & 11dB           & 12dB           & 13dB           & 14dB           &               \\
            \midrule
                                            & MM (base)                        & 16.53                                                                       & 27.39                                           & 41.47          & 58.52          & 75.18          & 88.15          & 0.99          \\
            u1                              & NTM                              & 25.38                                                                       & 39.92                                           & 58.38          & 75.96          & 90.04          & 96.93          & 0.99          \\
                                            & \textbf{MM+LCB}                  & \textbf{19.23}                                                              & \textbf{30.65}                                  & \textbf{46.19} & \textbf{63.70} & \textbf{80.16} & \textbf{91.68} & \textbf{1.00} \\
            \midrule
                                            & MM (base)                        & 23.83                                                                       & 37.88                                           & 55.58          & 73.82          & 88.13          & 96.38          & 0.99          \\
            u2                              & NTM                              & 28.23                                                                       & 44.19                                           & 63.89          & 81.56          & 93.82          & 98.53          & 0.94          \\
                                            & \textbf{MM+LCB}                  & \textbf{26.14}                                                              & \textbf{41.46}                                  & \textbf{60.29} & \textbf{78.12} & \textbf{91.44} & \textbf{97.64} & \textbf{0.96} \\
            \midrule
                                            & MM (base)                        & 24.76                                                                       & 39.20                                           & 57.68          & 75.48          & 89.34          & 96.83          & 0.99          \\
            u4                              & NTM                              & 29.10                                                                       & 45.15                                           & 64.89          & 82.22          & 94.03          & 98.58          & 0.97          \\
                                            & \textbf{MM+LCB}                  & \textbf{25.94}                                                              & \textbf{41.12}                                  & \textbf{60.14} & \textbf{77.89} & \textbf{91.49} & \textbf{97.68} & \textbf{0.96} \\
            \bottomrule
        \end{tabularx}}
\end{table}
\subsection{Strong Target Detection Performance}
This experiment assesses performance for $ \gamma>30 $, as shown in Table~\ref{t1l}. MM+LCB causes a minor decline in detection probability, but the loss within 2.5\% of MM is acceptable. Under these conditions, only NTM experiences overexposure, rendering the model nearly unusable.
\begin{table}[h] % aabbcc
    \renewcommand{\arraystretch}{1.} % 设置行高
    \caption{Results for strong SNR target}
    \label{t1l}
    \centering % 表格居中
    \setlength{\tabcolsep}{1.4mm}{
        % \fontsize{7.5}{8.5}\selectfont % 设置字体大小为 9pt, 行距为 11pt
        \begin{tabularx}{\linewidth}{c c c c c c c c} % 使用 tabularx, 设置总宽度为 \linewidth
            \toprule
            \multirow{2}{*}{\textbf{Model}} & \multirow{2}{*}{\textbf{Method}} & \multicolumn{5}{c}{$P_{\text{d}} \scriptstyle(\%)$ at different SNR levels} & \multirow{2}{*}{\makecell{$\bar{P}_{\text{fa}}$                                                                    \\ ($\scriptstyle\times10^{-4}$)}} \\
                                            &                                  & 30dB                                                                        & 40dB                                            & 50dB           & 60dB           & 70dB           &               \\
            \midrule
                                            & MM (base)                        & 99.95                                                                       & 99.85                                           & 98.64          & 97.33          & 96.86          & 1.88          \\
            u1                              & NTM                              & 8.19                                                                        & 0.19                                            & 0.05           & 0.05           & 0.05           & 70.65         \\
                                            & \textbf{MM+LCB}                  & \textbf{97.83}                                                              & \textbf{97.41}                                  & \textbf{97.45} & \textbf{96.67} & \textbf{94.56} & \textbf{1.84} \\
            \midrule
                                            & MM (base)                        & 100.00                                                                      & 99.99                                           & 100.00         & 99.98          & 99.74          & 1.85          \\
            u2                              & NTM                              & 26.14                                                                       & 0.33                                            & 0.10           & 0.05           & 0.06           & 72.28         \\
                                            & \textbf{MM+LCB}                  & \textbf{99.99}                                                              & \textbf{99.97}                                  & \textbf{99.79} & \textbf{99.76} & \textbf{99.70} & \textbf{1.25} \\
            \midrule
                                            & MM (base)                        & 99.86                                                                       & 99.87                                           & 99.95          & 100.00         & 100.00         & 1.51          \\
            u4                              & NTM                              & 65.20                                                                       & 35.53                                           & 23.17          & 18.27          & 17.76          & 29.13         \\
                                            & \textbf{MM+LCB}                  & \textbf{100.00}                                                             & \textbf{99.97}                                  & \textbf{99.92} & \textbf{99.86} & \textbf{99.76} & \textbf{1.30} \\
            \bottomrule
        \end{tabularx}}
\end{table}

\subsection{Urban Vehicle Detection Results}
\begin{figure}[h] % aabbcc
    \centering
    \includegraphics[width=1.\linewidth,page=6]{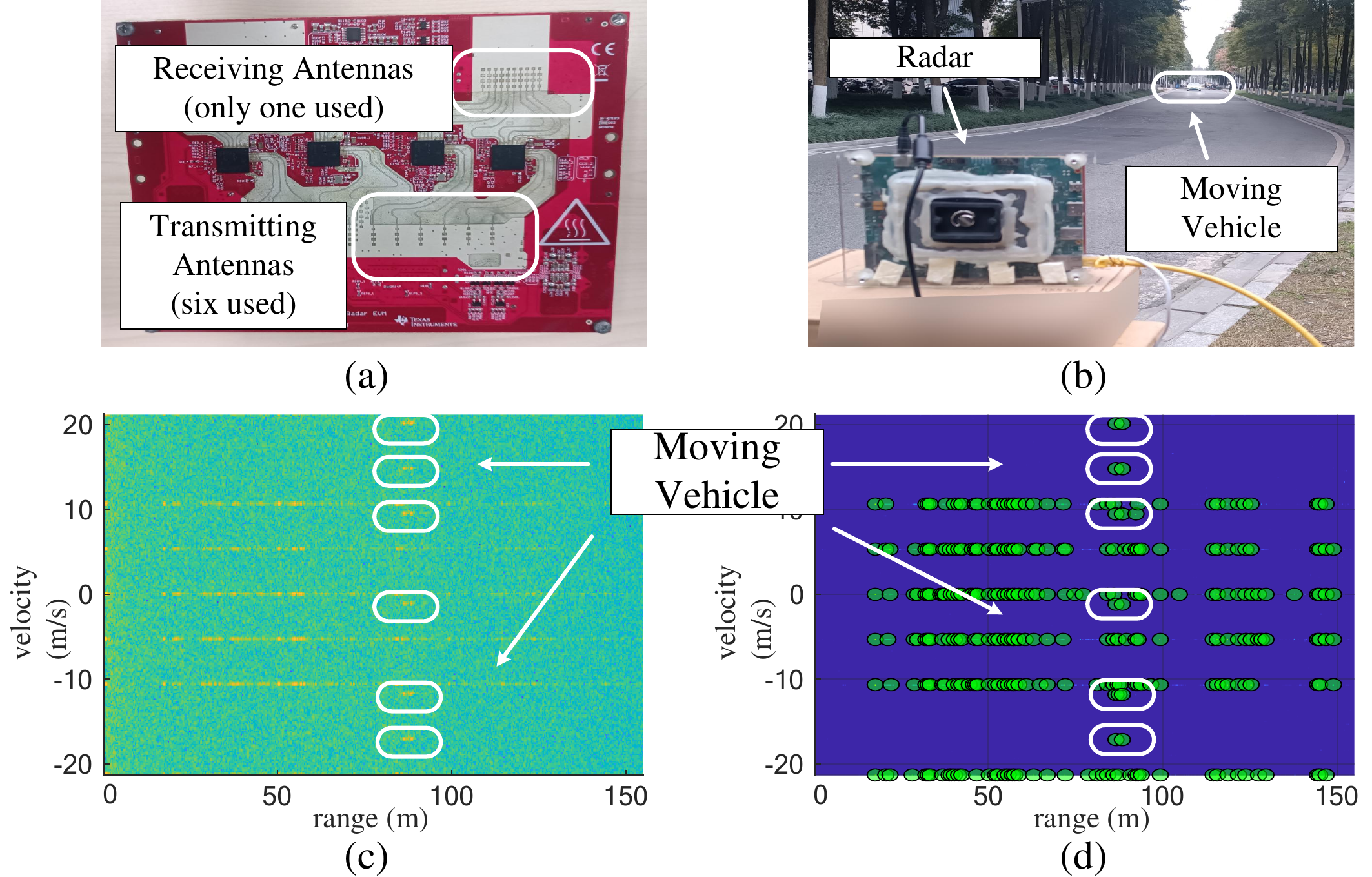}
    \vspace{-0.5cm}
    \caption{The detection result of real datasets. (a) radar; (b) the real scenario; (c) RDM; (d) the detection result.}
    \label{GT}
\end{figure}

We validate the proposed method using real datasets acquired from the TI AWR 2243 Cascade EVM radar, which is an embedded millimeter wave radar widely used in automotive, transportation, and drone applications. The radar is configured with identical parameters, and a two layer LCB-CV-UNet is trained using the dataset generated by the DHDC method. As shown in Fig.~\ref{GT}, the method effectively detects moving vehicles in the real scenarios. Additionally, a considerable number of static ground objects are detected along the zero velocity axis in the subbands with $ \alpha = 1\times10^{-6} $. This confirms that the proposed methods are effective on real signal data, adaptable to environments of high complexity, and capable of accurately detecting targets.
\vspace{-0.3cm}

\section{Conclusion}  % 扔掉一句话，不然arxiv版本就超页了，虽然似乎不影响什么
This paper presents a radar target detection method named LCB-CV-UNet, aimed at addressing the reduction in target detection probability caused by radar signals exhibiting HDR characteristics. Simulation experiments show certain advantages in target detection probability, and realistic scenario experiments validate the robustness of the model in complex environments. In summary, the proposed method is lightweight and efficient, offering a solution with superior performance for HDR signals in remote sensing applications, such as the edge-side radar target detection challenges.

\bibliographystyle{IEEEtranN}
\bibliography{references}

\end{document}